\documentclass[12pt,preprint]{aastex}
\usepackage[onecolumn]{emulateapj5}

\shorttitle{Parameters from SDSS Eigenmodes}
\shortauthors{Pope et al.}

\begin{document}

\title{Cosmological Parameters from Eigenmode Analysis of
Sloan Digital Sky Survey Galaxy Redshifts}

\author{
Adrian C. Pope\altaffilmark{1},
Takahiko Matsubara\altaffilmark{2},
Alexander S. Szalay\altaffilmark{1},
Michael R. Blanton\altaffilmark{3},
Daniel J. Eisenstein\altaffilmark{4},
Jim Gray\altaffilmark{5},
Bhuvnesh Jain\altaffilmark{6},
Neta A. Bahcall\altaffilmark{10},
Jon Brinkmann\altaffilmark{17},
Tamas Budavari\altaffilmark{1},
Andrew J. Connolly\altaffilmark{7},
Joshua A. Frieman\altaffilmark{8,9},
James E. Gunn\altaffilmark{10},
David Johnston\altaffilmark{8,9},
Stephen M. Kent\altaffilmark{8},
Robert H. Lupton\altaffilmark{10},
Avery Meiksin\altaffilmark{12},
Robert C. Nichol\altaffilmark{15},
Donald P. Schneider\altaffilmark{16},
Ryan Scranton\altaffilmark{7},
Michael A. Strauss\altaffilmark{10},
Istvan Szapudi\altaffilmark{13},
Max Tegmark\altaffilmark{6},
Michael S. Vogeley\altaffilmark{11},
David H. Weinberg\altaffilmark{14},
and Idit Zehavi\altaffilmark{9}
for the SDSS Collaboration}

\altaffiltext{1}{Johns Hopkins University}
\altaffiltext{2}{Nagoya University}
\altaffiltext{3}{New York University}
\altaffiltext{4}{University of Arizona}
\altaffiltext{5}{Microsoft Research}
\altaffiltext{6}{University of Pennsylvania}
\altaffiltext{7}{University of Pittsburgh}
\altaffiltext{8}{Fermi National Accelerator Laboratory}
\altaffiltext{9}{University of Chicago}
\altaffiltext{10}{Princeton University}
\altaffiltext{11}{Drexel University}
\altaffiltext{12}{University of Edinburgh}
\altaffiltext{13}{University of Hawaii}
\altaffiltext{14}{Ohio State University}
\altaffiltext{15}{Carnegie Mellon University}
\altaffiltext{16}{Pennsylvania State University}
\altaffiltext{17}{Apache Point Observatory}

\begin{abstract}

We present estimates of cosmological parameters from the application
of the Karhunen-Lo\`{e}ve transform to the analysis of the 3D power
spectrum of density fluctuations using Sloan Digital Sky Survey galaxy
redshifts.  We use $\Omega_mh$ and $f_b = \Omega_b/\Omega_m$ to
describe the shape of the power spectrum, $\sigma^L_{8g}$ for the
(linearly extrapolated) normalization, and $\beta$ to parametrize
linear theory redshift space distortions.   On scales $k \lesssim 0.16
h {\rm Mpc}^{-1}$, our maximum likelihood values are $\Omega_mh =
0.264 \pm 0.043$, $f_b = 0.286 \pm 0.065$, $\sigma^L_{8g} = 0.966 \pm
0.048$, and $\beta = 0.45 \pm 0.12$.  When we take a prior on
$\Omega_b$ from WMAP, we find $\Omega_mh = 0.207 \pm 0.030$, which is
in excellent agreement with WMAP and 2dF.  This indicates that we have
reasonably measured the gross shape of the power spectrum but we have
difficulty breaking the degeneracy between $\Omega_mh$ and $f_b$
because the baryon oscillations are not resolved in the current
spectroscopic survey window function.

\end{abstract}


\section{Introduction}

Redshift surveys are an extremely useful tool to study  the large
scale distribution of galaxies. Of the many possible statistical
estimators the power spectrum of the density fluctuations has emerged
as one of the easiest to connect to theories of structure formation in
the Universe, especially in the limit of Gaussian fluctuations where
the power spectrum is the complete statistical description.  There are
several ways to measure the power spectrum \citep[for a comparison of
techniques see][]{t1998}. Over the last few years, the
Karhunen-Lo\`{e}ve method \citep[, hereafter VS96]{vs1996} has
been recognized as the optimal way to build an orthogonal basis set
for likelihood analysis, even if the underlying survey has a very
irregular footprint on the sky.  A variant of the same technique is
used for the analysis of CMB fluctuations \citep{bond2000}.

The shape of the power spectrum is well described by a small set of
parameters \citep{eh1998}. For redshift surveys, it is of particular
importance to consider the large-scale anisotropies caused by infall
\citep{k1987}. Using a forward technique that compares models
directly to the data, like the KL-transform,
enables us to easily consider these anisotropies in full detail. Here
we present results of a parametric analysis of the shape of the
fluctuation spectrum for the SDSS galaxy catalog.

\section{Data}

\subsection{Sloan Digital Sky Survey}

The Sloan Digital Sky Survey \citep[SDSS;][]{y2000,sto2002} plans to
map nearly one quarter of the sky using a dedicated 2.5 meter
telescope at Apache Point Observatory in New Mexico.  A drift-scanning
CCD camera \citep{g1998} is used to image the sky with custom set of 5
filters ($ugriz$) \citep{f1996,smith2002} to a limiting
\citet{p1976} magnitude of $m_r \sim 22.5$.  Observations are
calibrated using a 0.5 meter photometric telescope \citep{pt2001}.
After a stripe of sky has been imaged, reduced, and astrometrically
calibrated \citep{astrom2003} , additional automated software selects
potential targets for spectroscopy.  These targets are assigned to
$3\degr$ diameter (possibly overlapping) circles on the sky called
tiles \citep{b2003tiling}.  Aluminum plates drilled from the tile
patterns hold optical fibers that feed into the SDSS spectrographs
\citep{u1999}.  The SDSS Main Galaxy Sample \citep[MGS;][]{str2002}
will consist of spectra of nearly one million low redshift ($\langle z
\rangle \sim 0.1$) galaxies creating a three dimensional map of local
large scale structure.

\subsection{Large Scale Structure Sample}
\label{sub:lss}

Considerable effort has been invested in preparing SDSS MGS redshift
data for large scale structure studies.  The first task is to correct
for fiber collisions.  The minimum separation between optical fibers
is $55 \arcsec$  which causes a correlated loss of redshifts in areas
covered by a single plate.  Galaxy targets that were not observed due to
collisions are assigned the redshift of their nearest neighbor.
Next the sky is divided into unique
regions of overlapping spectroscopic plates called sectors.  The angular
completeness is calculated for each sector as if the collided galaxies
had been successfully measured.  Galaxy magnitudes are
extinction-corrected with the \citet{sfd1998} dust maps, then
k-corrections are applied and rest frame colors and luminosities are
calculated \citep{b2003kcorr}.  Subsamples are created by making
appropriate cuts in luminosity, color, and/or flux.  A luminosity
function is then calculated for each subsample \citep{b2003lum} and
used to create a radial selection function assuming $\Omega_m = 0.3$
and $\Omega_\Lambda = 0.7$ cosmology.

This analysis considers two samples of SDSS data, which we will label
sample 10 and sample 12.  Both samples were prepared in similar manners,
although using different versions of software.  Sample 12 represents
a later state of the survey and the sample 10 area is contained in
sample 12.
Sample 10 represents 1983.39 completeness-weighted square
degrees of spectroscopically observed SDSS data and 165,812 
MGS redshifts.  Sample 12 has 205,484 redshifts over 2406.74
square degrees.
Both samples are larger than the 1360 square degrees
of spectroscopy in data release 1 \citep[DR1;][]{dr12003} of the SDSS.
The geometry of the samples and DR1 are qualitatively similar,
consisting of two thick slices in the northern cap of the survey and
three thin stripes in the south.  
The samples used have a luminosity cut of $-19 \geq M_r \geq -22$,
where $h = 1.0$ and $M_* = -20.44$ \citep{b2003lum}.
Rest frame quantities (ie absolute magnitudes) are given for the SDSS
filters at z=0.1, the median depth of the MGS.
In a study of the two point correlation function of SDSS galaxy redshifts,
\citet{zehavi2002} found that the bias relative to $M_*$ galaxies
varies from 0.8 for galaxies with $M = M_* + 1.5$ to 1.2 for galaxies
with $M = M_* - 1.5$.  \citet{norberg2001} found similar results for the
2dF, with the trend becoming more pronounced at luminosities significantly
greater than $L_*$.
The dependence of clustering strength on luminosity could induce an extra
tilt in the power spectrum because more luminous galaxies contribute more
at large scales and less luminous galaxies contribute more at small scales
due to the number of available baselines.  We minimize this effect by
staying within $M = M_* \pm  \sim 1.5$.
A uniform flux limit of $m_r \leq 17.5$ was applied, leaving
110,345 redshifts for sample 10 and 134,141 for sample 12.
Although there are luminosity limits for this sample, it is essentially
a flux limited sample with a (slowly) varying selection function.
We used galaxies in the redshift range $0.05 \le z \le 0.17$.

\section{Algorithm}

\subsection{The Karhunen-Lo\`{e}ve Eigenbasis}

Following the strategy described in VS96, the first step in
a Karhunen-Lo\`{e}ve (KL) eigenmode analysis of a redshift survey is
to divide the survey volume into cells and use the vector of galaxy
counts within the cells as our data.  This allows a large compression
in the size of the dataset without a loss of information on large
scales.  Our data vector of fluctuations $\bf d$ is defined as

\begin{equation}
d_i = c_i/n_i - 1
\label{eq:datavector}
\end{equation}

\noindent
where $c_i$ is the observed number of galaxies in the $i^{\rm th}$ cell and
$n_i = \langle c_i \rangle$ is its expected value, calculated from the
angular completeness and radial selection function.  The data is
``whitened'' by the factor $1/n_i$ to control shot noise properties in
the transform (VS96).  We call this the ``overdensity'' convention.

The KL modes are the solutions to the eigenvalue problem 
${\bf R} {\bf \Psi}_n
= \lambda_n {\bf \Psi}_n$
with the correlation matrix of the data given by

\begin{equation}
R_{ij} = \langle d_i d_j \rangle = 
\xi_{ij} + \delta_{ij}/n_i + \eta_{ij}/(n_in_j)
\label{eq:corrmatrix}
\end{equation}

\noindent
where $\xi_{ij}$ is the cell-averaged correlation matrix,
$\delta_{ij}/n_i$ is the shot noise term, and $\eta_{ij}/(n_i n_j)$
can be used to account for correlated noise (not used in this analysis).
The most obvious source of correlated noise in the MGS would be
differences in photometric zero points between different SDSS imaging
runs, which would result in ``zebra stripe'' patterns of density
fluctuations.  The MGS selection has a magnitude limit, but no color
selection terms, so the variation in target density depends only linearly
on the photometric calibration.  The $r$ band zero point variation is
0.02 mag rms \citep{dr12003}, indicating that the density variation
should be $\lesssim 2\%$.
The transformed data vector $\bf B$ is the expansion of
$\begin{bf}d\end{bf}$ over the KL modes ${\bf \Psi}_n$:

\begin{equation}
{\bf d} = \sum_n B_n {\bf \Psi}_n.
\label{eq:expansion}
\end{equation}

\noindent
The KL basis is defined by two properties: orthonormality of the
basis vectors, ${\bf \Psi}_m \cdot {\bf \Psi}_n
= \delta_{mn}$, and statistically orthogonality of the
transformed data, $\langle B_m B_n \rangle = \langle B_n^2 \rangle
\delta_{mn}$.

\subsection{The Correlation Function in Redshift Space}
\label{sub:rscf}

In order to directly compare cosmological models to our redshift data
using a two point statistic we must calculate the redshift space
correlation function
$\xi^{(s)} ({\bf r}_i,{\bf r}_j)$, where
${\bf r}_i$ and ${\bf r}_j$ 
describe positions in the observable angles and redshift.  The infall
onto large scale structures affects the velocities of galaxies leading
to an anisotropy in redshift space for a power spectrum that is
isotropic in real space \citep{k1987}.  \citet{sml1998} derived an
expansion of the correlation function that accounts for this
anisotropy in linear theory for arbitrary angles.  The expansion is

\begin{eqnarray}
&&
\xi^{(s)}({\bf r}_i,{\bf r}_j) =
c_{00}\xi_0^{(0)} + c_{02}\xi_2^{(0)} + c_{04}\xi_4^{(0)} + ...,
\label{eq:rscfexp} \\
&&
\xi_L^{(n)}(r) =  \case{1}{2\pi^2}\int dk k^2 k^{-n}j_L(kr)P(k)
\label{eq:rscfpart}
\end{eqnarray}

\noindent
where the $c_{nL}$ coefficients are polynomials of $\beta$ and
functions of the relative geometry of the two points.   The quantity
$\beta$ relates infall velocity to matter density and is well
approximated by the fitting formula  $\beta = \Omega_m^{0.6}/b$ where
$b$ is the bias parameter.   Further terms in Eq.~(\ref{eq:rscfexp})
are negligible as long as $2 + \partial {\rm ln} \phi(r) / \partial
{\rm ln} r$ (where $r$ is the  distance to the cell and $\phi(r)$ is
the radial selection function) does not significantly differ  (ie
orders of magnitude) from unity.  For the redshift range considered in
this analysis $|2 + \partial {\rm ln} \phi(r) / \partial {\rm ln} r|
\lesssim 4$.  When using counts-in-cells, we must calculate the
cell-averaged correlation matrix

\begin{equation}
\xi_{ij} = \int d^3{\bf r}_1 \int d^3{\bf r}_2
\xi^{(s)}({\bf r}_1,{\bf r}_2)
W_i({\bf x}_i - {\bf r}_1)
W_j({\bf x}_j - {\bf r}_2)
\label{eq:cellavgcm}
\end{equation}

\noindent
where $W_i({\bf y})$ is the cell window function and ${\bf x}_i$ is
the position of the $i^{\rm th}$ cell.  To be precise, $W_i({\bf y})$
should describe the shape of the cell in redshift space.  Numerical
calculation of this multi-dimensional integral can be computationally
expensive.  However, for the case of spherically symmetric cells we
can change the order of integration and perform the redshift space
integrals in Eq.~(\ref{eq:cellavgcm}) analytically before the
$k$-space integral in Eq.~(\ref{eq:rscfpart}).  If both cells have the
same window function, we can use Eq.~(\ref{eq:rscfexp}) as our
cell-averaged correlation function (with ${\bf r}_i$ and ${\bf r}_j$
indicating the cell positions) if we replace $P(k)$ with $P(k)
\tilde{W}^2(k)$ in Eq.~(\ref{eq:rscfpart}) where $\tilde{W}(k)$ is the
Fourier transform of the cell window function.  This results in a one
dimensional numerical integral.  The full technical details of our
method will be presented in  \citet[][in preparation]{matsu2004}.

We used hard spheres as our cell shape and placed them in a hexagonal
closest packed (the most efficient 3D packing, with a 74\%
space-filling factor) arrangement.  The current slice-like survey
geometry and packing arrangement causes some spheres to partially
protrude outside the survey.   The effective fraction of the sphere
that is sampled is also affected by the angular completeness  of our
survey (which averages $\sim 97\%$).  We calculate our expected counts
as if the sphere was entirely filled and multiply the observed galaxy
counts by  $1/f_i$ where $f_i$ is the fraction of the $i$th sphere's
volume that was effectively sampled.  This sparser sampling also
increases the shot noise by a factor of $1/f_i$.  Cells with $f_i <
0.65$ were rejected as too incomplete.  We found that a $6 h^{-1}{\rm
Mpc}$ sphere radius allowed us to fill the survey volume with a
computationally feasible number of cells without the spheres
protruding too much out of the survey, while smoothing on sufficiently
small length scales so that we do not lose information in the linear
regime ($2\pi/k \gtrsim 40 h^{-1}{\rm Mpc}$).  We used 14,194 cells for 
sample 10 and 16,924 for sample 12.

The calculation of the sampling fraction for each cell is difficult due
to the complicated shapes of the sectors (\S \ref{sub:lss}).
We created a high resolution angular completeness map in a SQL Server
database using $10^7$ random angular points over the entire sky.
Each point was assigned a completeness weighting by finding which sector 
contained the point or setting the completeness to zero for points outside
the survey area.
We used a Hierarchical Triangular Mesh \citep[HTM;][]{kunszt2001} spatial
indexing scheme to find all points in the completeness map that pierce a
cell and calculate the volume weighted completeness for that cell.

\subsection{Eigenmode Selection}
\label{sub:modeSelection}

The KL transform is linear, so there is no loss of information if we
use all of the eigenmodes.  However, if we perform a truncated
expansion we can use the KL transform for compression and filtering.
The difference between the original data vector and a 
truncated reconstruction,
${\bf \hat{d}} = \sum_{i=1}^{M<N} B_i {\bf \Psi}_i$,
where we use only $M$ out of a possible $N$ modes can be related to the
eigenvalues of the excluded modes by
$({\bf d} - {\bf \hat{d}})^2 = \sum_{i=M+1}^N \lambda_i.$
The error is minimized (in a squared sense) when we retain modes with
larger eigenvalues and drop modes with smaller eigenvalues, which is
sometimes called optimal subspace filtering \citep{t1992}.

The eigenvalue of a KL mode is also related to the range in $k$-space
sampled by that mode.  Our models assume that linear theory is a good
approximation, which is only valid on larger scales.  Consequently we
only wish to use KL modes that fall inside a ``Fermi sphere'' whose
radius is set by our cutoff wavenumber $k_f$.  If we sort modes by
decreasing eigenvalue, they will densely pack $k$-space starting from
the origin.  The modes resist overlapping in $k$-space due to
orthogonality.  The shape of a KL mode in $k$-space resembles the
Fourier transform of the survey window function.  This means that the
number of KL modes within the ``Fermi sphere'' depends mostly on the
survey window function and does not drastically change if we change
the size of our cells, as long as we have significantly more cells
than modes (which means that our cells must be smaller than the cutoff
wavelength).  In a fully three dimensional survey the modes would fill
$k$-space roughly spherically and $M \propto k_f^3$.  However, the
current SDSS geometry resembles several two dimensional slices,
resulting in KL modes that resemble cigars in $k$-space.  These modes
pack layer-by-layer into spherical shells whose diameters are integer
multiples of the long axis of the mode.  See Fig. 5 in \citet{s2003}
for a visualization.  This results in a scaling more like $M \propto
k_f^2$.

\begin{figure}
\plotone{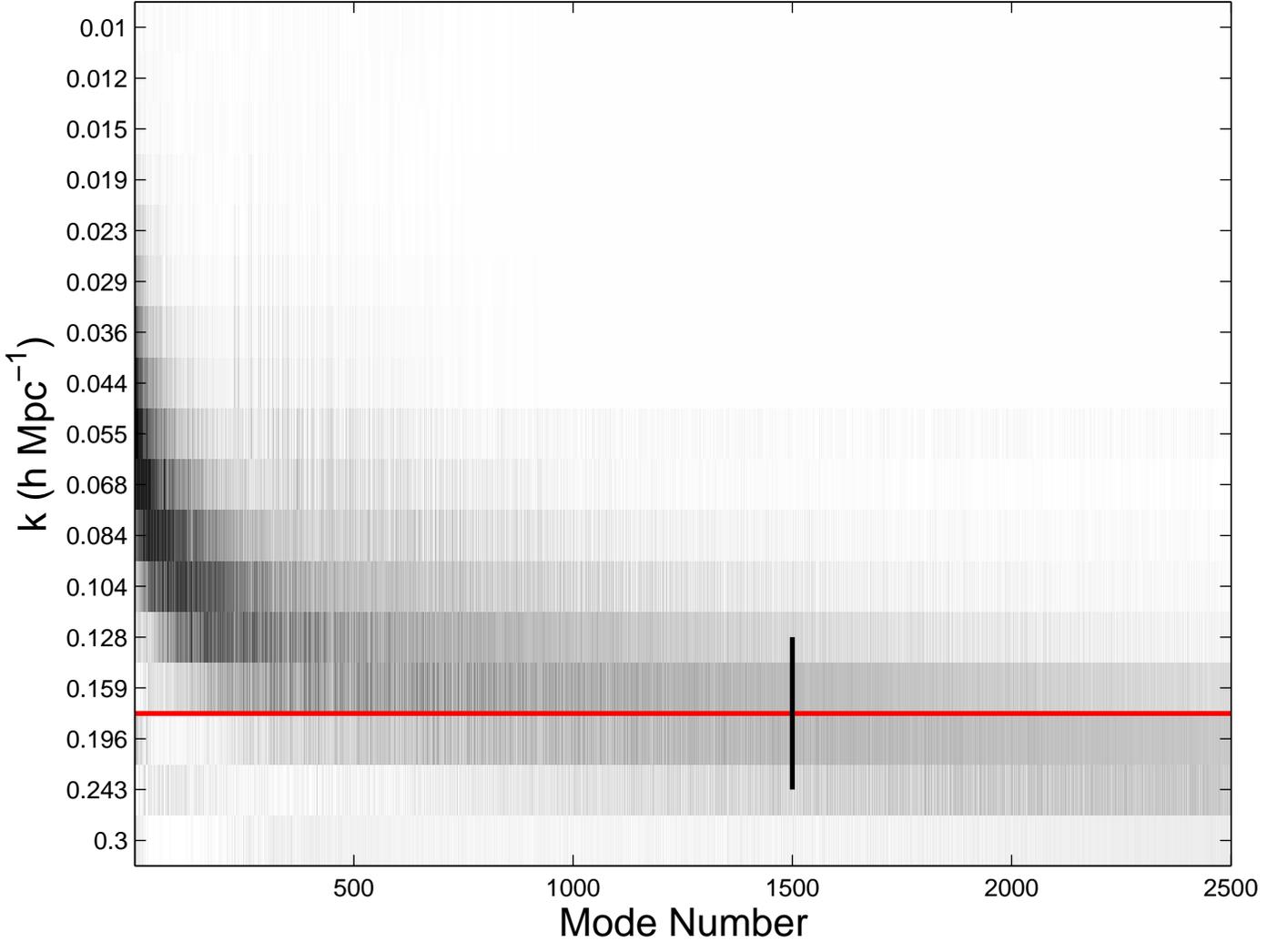}
\caption{
Grayscale image of wave number vs. mode number.  The  horizontal red
line indicates $k_f = 0.16 h {\rm Mpc}^{-1}$.  The vertical black line
indicates the truncated number of modes used for likelihood analysis.
}
\label{fig:modes}
\end{figure}

In choosing the number of KL modes to use in our analysis we try to
keep as many modes as possible for better constraints on our parameter
values while requiring that our modes are consistent with linear
theory.  We have developed a convenient method for determing the range
in $k$-space probed by each KL mode.  We separate the integral in
Eq.~(\ref{eq:rscfpart}) into bandpowers in $k$.  This allows us to
determine how strongly each mode couples to each bandpower, which
shows a coarse picture of the spherically averaged position of the
mode in $k$-space.  Fig. \ref{fig:modes} shows a grayscale image of
how the modes couple to the bandpowers.  Once we choose a value for the
cutoff wavenumber $k_f$, we truncate our expansion at the mode where 
wavenumbers larger than $k_f$ start to dominate.

We can use the statistical properties of the transformed data to check
that we are avoiding non-linearities.  A rescaled version of the KL
coefficients $b_n = B_n / \sqrt{{\bf \Psi}_n}$ should be
normally distributed.  Non-linear effects would cause skewness (third
moment) and/or kurtosis (fourth moment) in the  distribution of $b_n$.
We do not see evidence of non-linear effects when we use $k_f \lesssim
0.16 h {\rm Mpc}^{-1}$ (corresponding to length scales $2 \pi/k_f \gtrsim
40 h^{-1}{\rm Mpc}$).
This value for the cutoff wavenumber leaves
us with 1500 modes for sample 10 and 1850 modes for sample 12.

\subsection{Model Testing}
\label{sub:modeltest}

We estimate cosmological parameters by performing maximum likelihood
analysis in KL space.  The likelihood of the observed data given a
model $m$ is

\begin{equation}
{\cal L}({\bf B}|m) = (2\pi)^{-M/2} 
\vert {\bf C}_{m}\vert^{-1/2}
\exp\left[-\case{1}{2}{\bf B}^T
{\bf C}^{-1}_{m}
{\bf B}\right]
\label{eq:lh}
\end{equation}

\noindent
where ${\bf C}_{m}$ is the covariance matrix and can
be calculated as the projected model correlation matrix,

\begin{equation}
\left({\bf C}_{m}\right)_{ij} =
\langle B_i B_j \rangle_{m} =
{\bf \Psi}^T_i {\bf R}_{m} {\bf \Psi}_j.
\label{eq:proj}
\end{equation}

\noindent
Our method is based upon a linear comparison of models to data, thus
the ${\bf R}_m$ (and ${\bf C}_m$) model matrices only contain second
moments of the density field.  This linear estimator is
computationally more expensive than quadratic or higher order
estimators, but the results are less sensitive to non-linearities.
For a comparison of different estimation methods, see \citet{t1998}.

In practice we must decide on an explicit parametrization.  We
construct a power spectrum assuming a primordial spectrum of
fluctuations with a spectral index $n_s = 1$.  We use a fitting
formula from~\citet{eh1998} to characterize the transfer function,
including the baryon oscillations.  We fit for $\Omega_mh$ and $f_b =
\Omega_b/\Omega_m$ while taking a prior of $H_0 = 72 \pm 8~{\rm
km~s}^{-1}$ from the Hubble key project \citep{f2001} and fixing
$T_{\rm CMB} = 2.728 K$ \citep{fix1996}.  We fit the linearly
extrapolated $\sigma^L_{8g}$ for normalization, where $\sigma^L_{8g} =
b \sigma_{8m}$ and $b$ is the bias.  Linear theory redshift-space
distortions are characterized by $\beta$ (see \S \ref{sub:rscf}).

In order to search an appreciable portion of parameter space we
have developped efficient methods to calculate the model covariance
matrices ${\bf C}_{m}$.  The straightforward approach would
be to calculate the model correlation matrix for a set of parameters
and then project into the KL basis and calculate a likelihood, but
this is computationally expensive.  The covariance matrix can easily
be written as a linear combination of matrices and powers of
$\sigma^L_{8g}$ and $\beta$ (see \S \ref{sub:rscf}), so we can
project pieces of the correlation matrix and add them in the
appropriate proportions for those parameters.  However, the shape of
the power spectrum depends on $\Omega_m$, $f_b$, and $H_0$ in a
non-trivial way.  We project each bandpower of the correlation matrix
(see \S \ref{sub:modeSelection}) separately and add the pieces of
the covariance matrix together with appropriate weighting to represent
different power spectrum shapes.  This alleviates the need for further
projections.  We must be careful when choosing our bandpowers so that
we retain sufficient resolution to accurately mimic power spectrum
shapes (especially baryon oscillations), but we must also be careful
that our $k$ ranges are large enough that the integrals converge
correctly.

Note that a non-optimal choice of fiducial parameters does not bias our
results, but it can result in non-minimal error bars.  This procedure
can be iterated if necessary.

\section{Results and Discussion}

Our best-fit maximum-likelihood parameter values for samples 10 and 12
are presented in Table \ref{tab:results}.  Results are given for the
priors described in \S \ref{sub:modeltest} and also when using
the additional prior $\Omega_b = 0.047 \pm 0.006$ from WMAP
\citep{spergel2003}.
We show the results of sample 10 and 12 to give some indication of
sample variance, although sample 10 is a subset of sample 12.

\begin{figure}
\plotone{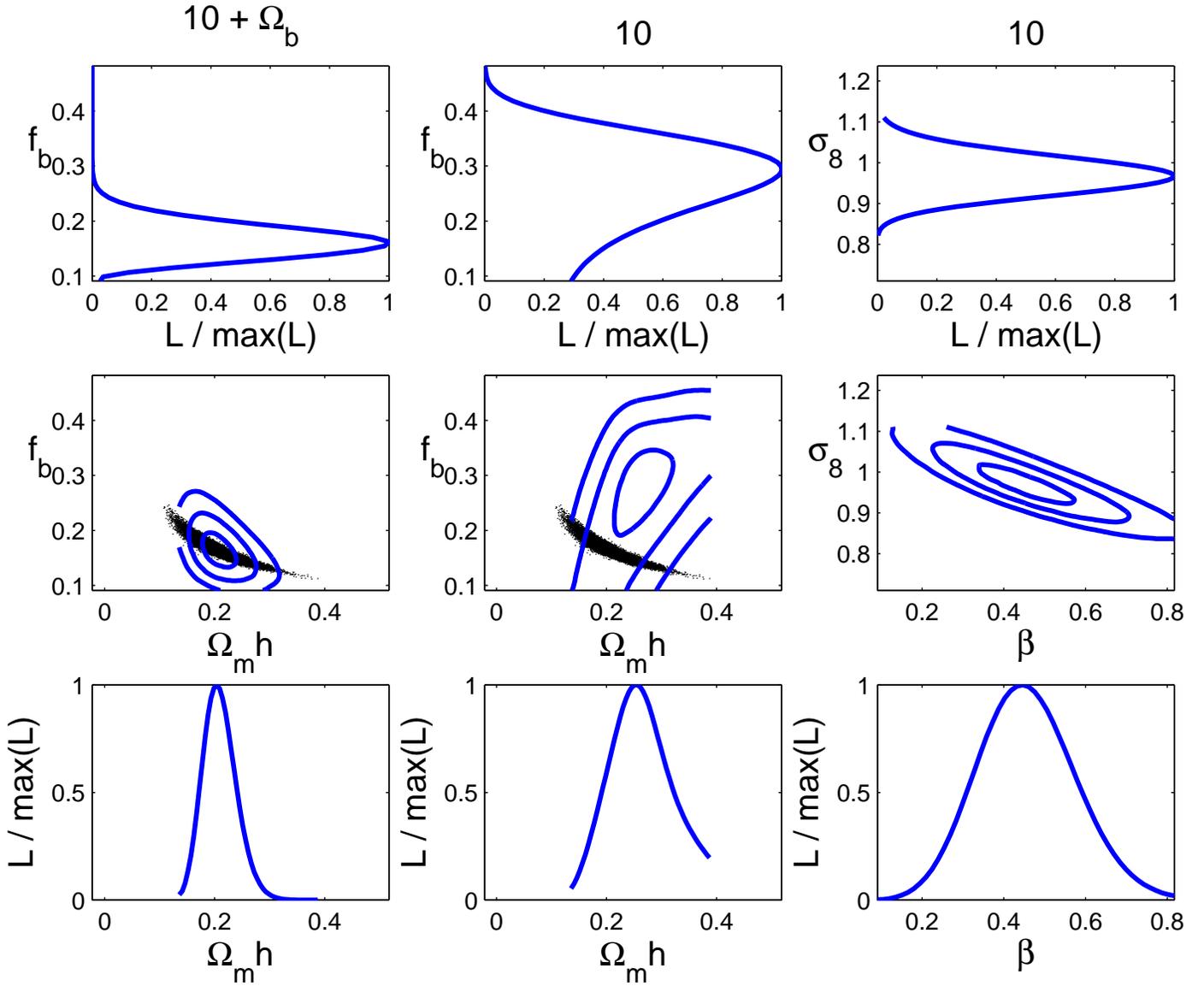}
\caption{
Likelihoods for parameters using sample 10.
The left column shows the power spectrum shape parameters with an 
$\Omega_b$ prior.
The middle column shows the power spectrum shape parameters without an
$\Omega_b$ prior.  
The right column shows normalization and distortion parameters.
The contours in the joint parameter plots are the two-dimensional 
1, 2, and 3 $\sigma$ contours.
The points in the $f_b$ vs. $\Omega_mh$ plots are MCMC points from WMAP
(alone).
Parameter combinations not plotted are nearly uncorrelated.
}
\label{fig:results}
\end{figure}

\begin{figure}
\plotone{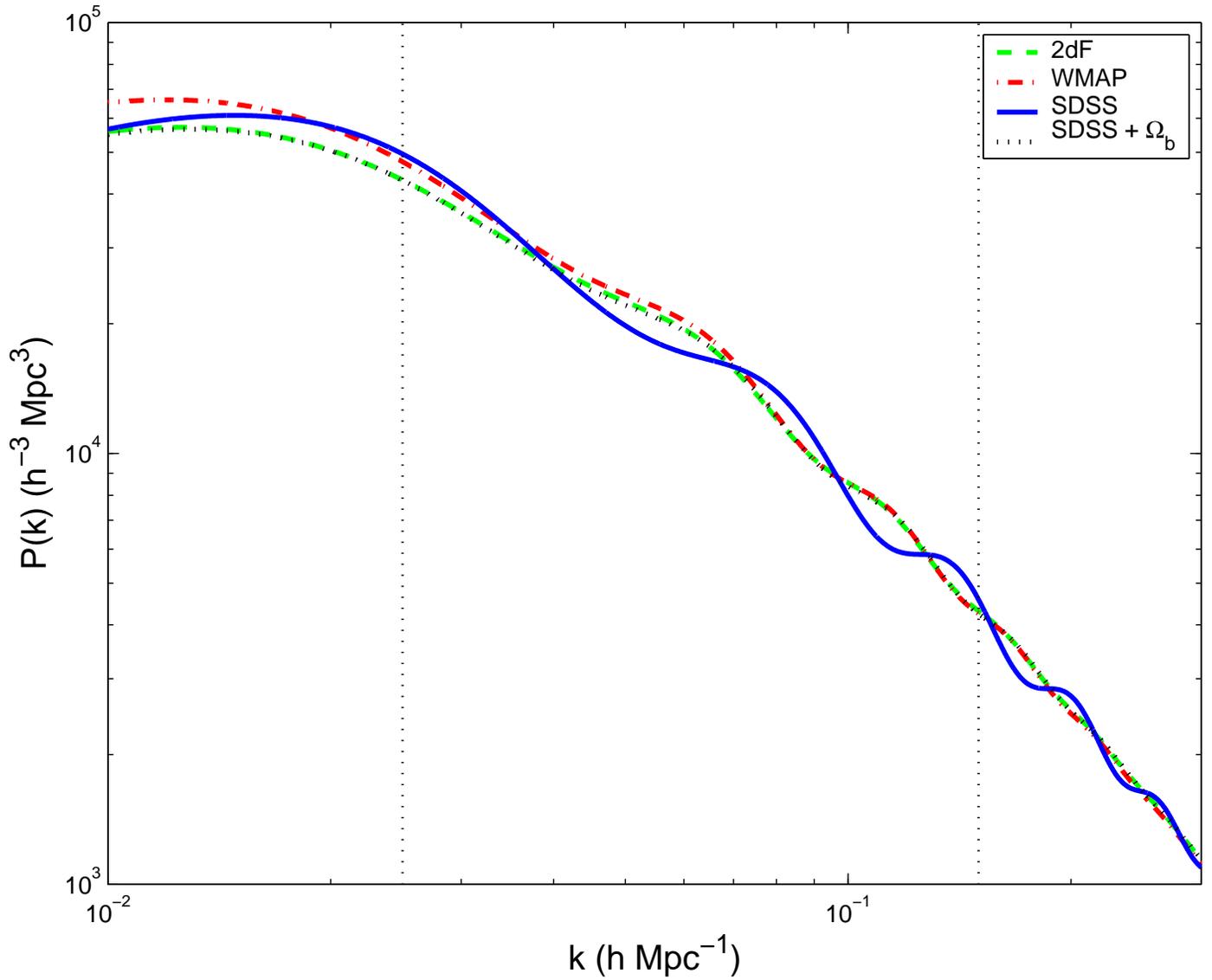}
\caption{
Plots of the real space P(k) from best-fit model parameters for 
SDSS (sample 10 with and without the $\Omega_b$ prior), WMAP, and 2dF.
All use $\sigma^L_{8g}$ from the SDSS for normalization.
The vertical dotted lines indicate the range in $k$ used in the SDSS analysis.
}
\label{fig:pk}
\end{figure}

The middle column of Fig.~\ref{fig:results} shows the marginalized
one-dimensional and two-dimensional confidence regions for the power
spectrum shape parameters $\Omega_mh$ and $f_b$ for sample 10 without
the additional prior on $\Omega_b$.  There is a strong correlation
between $\Omega_mh$ and $f_b$.  The gross shape of the power spectrum
(ie ignoring the baryon oscillations and concentrating on the position
of the peak and slope of the tail) is nearly constant along the ridge
of this correlation due to a degeneracy between shifting the position
of the peak with $\Omega_mh$ and adding power to the peak with $f_b$.
However, the strength of the baryon oscillations varies significantly
over this range.  Table \ref{tab:results} shows that our estimates of
$\Omega_mh$ agree well with the WMAP value of $0.194 \pm 0.04$
\citep{spergel2003} and the 2dF value of $0.20 \pm 0.03$
\citep{percival2001} when we use the additional prior on $\Omega_b$,
and the associated confidence regions are shown in the left column of
Fig.~\ref{fig:results}.  The results with the $\Omega_b$ prior
indicate that the gross shape of the power spectrum we measure is
consistent with WMAP and 2dF, as can be seen in Fig.~\ref{fig:pk} which
shows the (isotropic) real-space power spectra inferred from the
cosmological parameter estimates from the three surveys.
However, the results without the
$\Omega_b$ prior show that we have difficulty breaking the degeneracy
between $\Omega_mh$ and $f_b$ because the baryon oscillations are not
resolved due to the current state of the SDSS window function.

The right column of Fig.~\ref{fig:results} shows the marginalized
one-dimensional and two-dimensional confidence regions for
$\sigma^L_{8g}$ (normalization) and $\beta$ (distortions) for sample
10.  Again there is a strong correlation between these parameters,
which is expected from their dependence on $b$.  Our constraint on
$\sigma^L_{8g}$ is strong, but we can only measure $\beta$ to $\sim 20
\%$ which limits our ability to perform an independent estimate of
$b$.  We can compare our results to WMAP by examining the combination
of parameters $\sigma^L_{8g}\beta = \sigma^L_{8m}\Omega_m^{0.6}$, for
which we obtain the value $0.44 \pm 0.12$, in excellent agreement with
the WMAP result of $0.44 \pm 0.10$ \citep{spergel2003}.  By combining
our measurements with WMAP results we find $b = 1.07 \pm 0.13$ for our
galaxy sample, but this compares information dominated by galaxies
with redshifts $0.1 \lesssim z \lesssim 0.15$ to present-day matter.
If we use a $\Lambda CDM$ model to extrapolate to the present, we
would find $b \approx 1.16$.  Our galaxies cover a range of
luminosities but our signal is dominated by the more luminous galaxies
(brighter than $L_*$) because there are more long baselines available
for the more distant galaxies.  This must be kept in mind when comparing
our measurement of $\sigma^L_{8g}$ with other estimates using SDSS
data which focus on $L_*$ galaxies \citep{s2003,tegmark2003}

This analysis used less than one third of the data that will comprise
the completed SDSS survey.  Our ability to measure cosmological parameters
will increase as the survey area increases, but we should also gain
leverage in resolving features in the power spectrum
as our survey window function becomes cleaner.  The thickest slice of
data from the samples used was roughly $10\degr$, implying a
thickness of $\sim 50 h^{-1}{\rm Mpc}$ at $z \sim 0.1$.  As the slices
become thicker, the KL modes will become much more compact in that
direction in $k$-space.  Thus we will benefit from the change in the
survey aspect ratio in addition to the increase in survey area.

\acknowledgements

We would like to thank HP/Compaq for donating several Intel Itanium
large memory workstations used for this project.
This material is based upon work supported under a National Science 
Foundation Graduate Research Fellowship and by NSF AST-9802980 at
Johns Hopkins University.

Funding for the creation and distribution of the SDSS Archive has been
provided by the Alfred P. Sloan Foundation, the Participating
Institutions, the National Aeronautics and Space Administration, the
National Science Foundation, the U.S. Department of Energy, the
Japanese Monbukagakusho, and the Max Planck Society. The SDSS Web site
is http://www.sdss.org/.

The SDSS is managed by the Astrophysical Research Consortium (ARC) for
the Participating Institutions. The Participating Institutions are The
University of Chicago, Fermilab, the Institute for Advanced Study, the
Japan Participation Group, The Johns Hopkins University, Los Alamos
National Laboratory, the Max-Planck-Institute for Astronomy (MPIA),
the Max-Planck-Institute for Astrophysics (MPA), New Mexico State
University, University of Pittsburgh, Princeton University, the United
States Naval Observatory, and the University of Washington.

\begin{deluxetable}{lcccc}
\tablecaption{
Best Fit Parameter Values
\label{tab:results}}
\tablewidth{0pt}
\tablehead{
\colhead{} & \colhead{$10$} & \colhead{$10 + \Omega_b$} & \colhead{$12$} & \colhead{$12 + \Omega_b$}}
\startdata
$\Omega h$ & $0.264 \pm 0.043$ & $0.207 \pm 0.030$ & $0.270 \pm 0.057$ & $0.229 \pm 0.029$ \\
$f_b$ & $0.286 \pm 0.065$ & $0.163 \pm 0.031$ & $0.233 \pm 0.088$ & $0.149 \pm 0.026$ \\
$\sigma^L_{8g}$ & $0.966 \pm 0.048$ & $0.971 \pm 0.049$ & $0.978 \pm 0.043$ & $0.980 \pm 0.043$ \\
$\beta$ & $0.45 \pm 0.12$ & $0.44 \pm 0.12$ & $0.44 \pm 0.11$ & $0.43 \pm 0.11$ \\
\enddata
\tablecomments{
Maximum likelihood parameter values and 68\% confidences (marginalized
over all other parameters).  $\Omega_b$ indicates that a WMAP prior
was used.
}
\end{deluxetable}

\end{document}